\documentclass[twocolumn]{revtex4}

\usepackage{epsfig}

\begin{document}

\title{Molecular dynamics of shock fronts and their transitions in Lennard-Jonesium and Tin}

\author{J. M. D. Lane}
\email{mlane@chaos.ph.utexas.edu}
\affiliation{Center for Nonlinear Dynamics, University of Texas at Austin}

\author{M. P. Marder}
\email{marder@chaos.ph.utexas.edu}
\affiliation{Center for Nonlinear Dynamics, University of Texas at Austin}

\begin{abstract}
We develop a Continuous Hugoniot Method for the efficient simulation of shock wave fronts with molecular dynamics.  This  approach achieves a significantly improved efficiency in the generation of a dense sampling of steady-state shock front states, and allows for the study of shocks as a function of a continuous shock strength parameter, $v_p$.  This is, to our knowledge, the first attempt to map out the Hugoniot in a continuous fashion.

We first apply this method to shocks in single-crystal Lennard-Jonesium along the $\langle100\rangle$ direction.  Excellent agreement is found with both the published Lennard-Jones Hugoniot and results of conventional simulation methods.

We next present a continuous numerical Hugoniot for shocks in tin which agrees to within 6\% with experimental data.  We study the strong shock to elastic-plastic shock transition in tin and find that it is a continuous transition consistent with a transcritical bifurcation.
\end{abstract}

\date{\today}

\maketitle

\section{Introduction}
Laser-induced shock waves and femtosecond time-resolved optical diagnostics are providing a new window into shock research.  Recent shock physics experiments have made atomic response measurements at length scales and time scales well matched with computational molecular dynamics.  There are new opportunities for direct collaboration between experiment and simulation in the study of dynamics at the shock front.
 
This work presents a method which allows effective collaboration with experiment in characterizing the dynamics and melt properties at the shock front in real materials.  Recent experiments have indicated that the dynamic response at the shock front in solids can depend strongly not only on shock strength but also on failure or transition timescales \cite{loveridge-smith.01}.  On-going experiments at the High-Intensity Laser Science Group at the University of Texas at Austin are investigating the shock-strength dependence of melt time scales in tin and aluminum using table-top and larger laser systems in the terawatt range.

There are two major difficulties in applying conventional simulation methods to the study of dynamics near a shock front:  (1) To produce the environment at the front, one must simulate a large and ever-growing system, of which the front constitutes only a very small fraction; and (2) The conditions within a steady-state shock take long times to arise, and each computationally-expensive shock run results in only a single data point.

Our goal in this paper is to address each of these deficiencies with an efficient approach.  Our method, introduced in Section II, focuses computational resources only on the shock front, while simultaneously producing a continuum of shock strength final states in a single run.

The constrained dynamics methods of the Hugoniostat and others \cite{maillet.00, reed.02, reed.03} offer solutions to the first problem, but offer no information about the non-equilibrium dynamics at the shock front.  The approach of Zhakhovskii et al. \cite{zhakhovskii.99} succeeds in addressing the first point at the shock front, but does not address the second.  We generalize and expand on these methods here.  First, we concentrate our efforts on the neighborhood of the shock interface, thereby increasing computational efficiency, and second, we map system response to a continuum of shock strength final states by continuously varying the parameter within the simulation.  This combination of techniques we call the Continuous Hugoniot Method.

The shock Hugoniot is a relationship between any two variables of the final material state behind a shock.  It is derived from the conservation of mass, momentum, and energy across a shock front and the equation of state (EOS) for the material.  The conservation equations are
\begin{eqnarray}
 \frac{\rho_o}{\rho} & = & \frac{V}{V_o} \ = \ 1 - \frac{u_p}{U_S} \\ \nonumber\\
 P_{xx} & = & U_S\,u_p\,\rho_o \\ \nonumber\\
 E - E_o & = & \frac{1}{2}P_{xx}(V_o - V) \label{e:energy_jump}
\end{eqnarray}
where $U$, $u$, $\rho=\frac{1}{V}$, $P$ and $E$, are the shock velocity, particle velocity, density, pressure tensor and internal energy, respectively.   The subscript $0$ indicates the initial material state ahead of the shock front. The state variables refer to values far behind the shock front. $P_{xx}$ refers to the  component of the pressure tensor far behind the shock front perpendicular to the direction of the front.  The EOS is material specific.  The Hugoniot, a curve derived from these four equations and five unknowns, is a locus of final states.  Each point on the Hugoniot curve represents the final state of an individual shock experiment or simulation.

\section{Continuous Hugoniot Method}
Introductions to shock physics \cite{ahrens.92, graham.92} state clearly that the Hugoniot is not a thermodynamic path.  However, our simulation method makes the Hugoniot the thermodynamic path of our simulation system.  This is not thought to be possible experimentally.  Figure \ref{f:ourpath_hugoniot} shows (A) the experimental loading paths (Rayleigh lines) to each state of the Hugoniot and (B) the loading path which we will use to reach the same state points.  The following was briefly summarized in \cite{lane.06}.

\begin{figure}[htb]
\epsfig{file=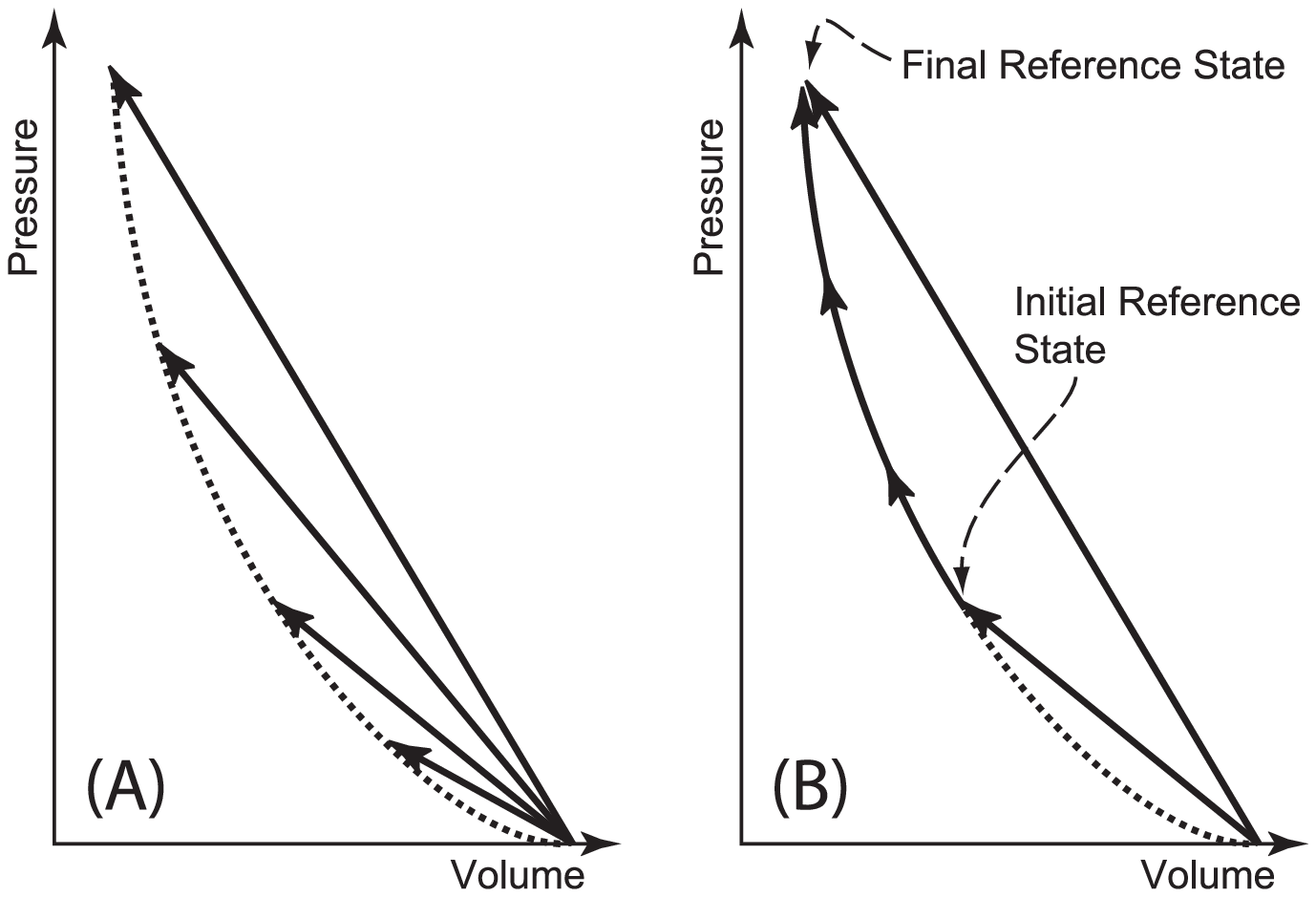,width=3.3in}
\caption{{\bf Hugoniot as a destination versus path} -- (A) The Hugoniot (dotted line) is illustrated as a collection of final shock states.  (B) The Continuous Hugoniot Method follows the path of the Hugoniot continuously from the initial reference state.  The solid lines in both plots depict the path of the system through $P$-$V$ space.}
\label{f:ourpath_hugoniot}
\end{figure}

\vspace{0.25in}
\noindent {\bf Conventional Simulation Reference Runs}\\
The Continuous Hugoniot Method results in a continuum of shock states which range from some initial shock strength to some final shock strength.  Two conventional molecular dynamics shock wave computations bookend this continuum.  Conventional simulations are runs which are driven by a warm impactor (momentum mirror) and are allowed to evolve for long times until they converge to a steady state.  These reference systems generally grow very large and are therefore computationally expensive.  Runs of this type are, however, necessary to seed our method and bound its error.

\begin{figure*}
\epsfig{file=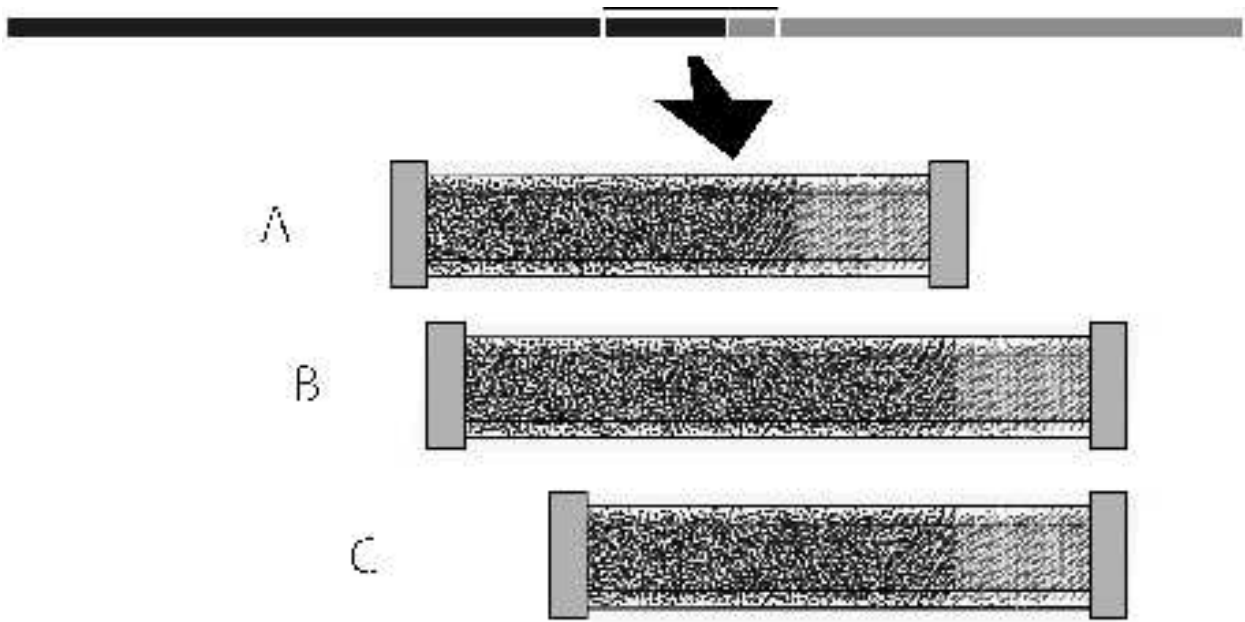,width=6.5in}
\caption{{\bf Co-moving frame centered on the shock front} -- From an initial reference state (A) a reduced system of particles are extracted.  A thermostated piston drives the system from the left and particles on the right are frozen.  (B) The system evolves as the piston drives forward.  As the shock front advances, fresh material is added to the right.  (C) When the material at the left has thermalized ahead of the piston, a block of material is purged and the cycle repeats.}
\label{f:ourpath_hugoniot}
\end{figure*}

The method begins with a conventional run which leaves the system in the initial Hugoniot reference state.  This run is illustrated in Fig. \ref{f:ourpath_hugoniot} (B) by the straight line chord (Rayleigh Line) connecting the initial condition to the Hugoniot.  This reference run provides a initial Hugoniot state.  We determine the smallest system necessary to model the shock front from this state.  The reduced system size is found by measuring the distance behind the shock front at which the material has thermalized.  Thermalization is determined from velocity distribution analysis at increasing distances from the shock front.  For our purposes, we consider the system thermalized when the distributions in the transverse and propogation directions are Maxwell-Boltzmann and give the same temperatures.

\vspace{0.25in}
\noindent {\bf System Reduction}\\
We truncate the reference system to a fixed width, leaving pristine material ahead and slab of compressed material behind.  The reduced system width is determined by the reference state thermalization length we have measured.  This reduced system is then evolved with material added and purged based on criterion established to keep the shock front in the center of the fixed-width sample.  At the rear purge point, we preserve the velocity distributions with a Langevin thermostat, and a momentum mirror prevents particle loss.  The piston velocity and mean velocity of the thermostat are equal, and this value, $v_p$, serves as the external control parameter for shock strength.

A buffer zone of undisturbed crystal must always exist ahead of the shock front, but this buffer need not be large. The shock travels faster than any wave in the uncompressed material, so we do not need to be concerned with preheating or dispersion. The buffer zone we maintain is 5 to 10 lattice planes wide. The initial temperature is set by another Langevin thermostat, and particles within a unit cell of the system's forward edge are frozen in place to prevent unphysical surface reconstructions.

As the system evolves, the shock front consumes material and advances into the buffer zone. The shock front location is continually updated at a set interval of timesteps.  The front is determined by analysis of the gradient in total energy per particle along the propogation direction.  The front location is used to trigger an advance event when the size of the buffer zone (measured from the system's forward edge to the shock front) falls below the user defined minimum.

At the advance event, a block of crystal is generated and appended to the forward system boundary. The generated crystal must have thickness equal to an integer multiple of the crystal unit cell and have zero initial temperature to prevent a mismatch with the frozen forward edge of the system.

The requirements for a smooth and valid purge event will be discussed at end of this section. For now, we assume the requirements are satisfied and discuss the steps of the purge process.  At each advance event, the total system size is calculated. A purge is initiated if the system size exceeds a user-defined maximum. The truncation of the system behind the front can be very delicate because the system is driven from behind. Care must be taken to preserve the thermodynamic properties at the back edge of the system and not to introduce any unphysical disturbance that might propagate.

Material is purged in a block equal in volume to the block appended during an advance. Because the purged material is compressed, however, there are more particles purged than appended in a single event. For this reason, there are fewer purge events than advance events.  A purge begins with a temperature measurement at the purge point. The back boundary is then shifted forward and particles beyond the new boundary are dropped from the system arrays. From this point, the system is evolved normally.

Our goals in this system reduction method are to simulate without approximation the dynamics at the front of a shock wave without devoting computing resources to simulate the entire system behind the shock. To do this, we require (1) that the final shock be in a steady state, and (2) that a thermalized equilibrium thermodynamic state evolve quickly (i.e. within a short distance) behind the shock front.  In some cases, this thermodynamic state is the equilibrium final state given by the Hugoniot relation. In others, it is an intermediate state. Our method requires only that we can replicate the thermodynamic properties at the purge with a Maxwell-Boltzmann velocity distribution.

It was observed early in this work that the piston boundary conditions play a very important role in the ability to drive the system smoothly.  It must be remembered here that the shockwave is supersonic in the uncompressed material, but subsonic in the compressed material behind the front.  This means that disturbances from the piston travel faster than the shock front and will eventually interact with it.  In traditional shock simulations, one does not need to worry about the piston interaction because as the piston grows further and further isolated from the shock front the two become decoupled and interact only in an average thermodynamic way.  In these large systems, any piston-generated spatial correlations in the system decay away in a trivially small percentage of the whole system.

\begin{figure}[htb]
\epsfig{file=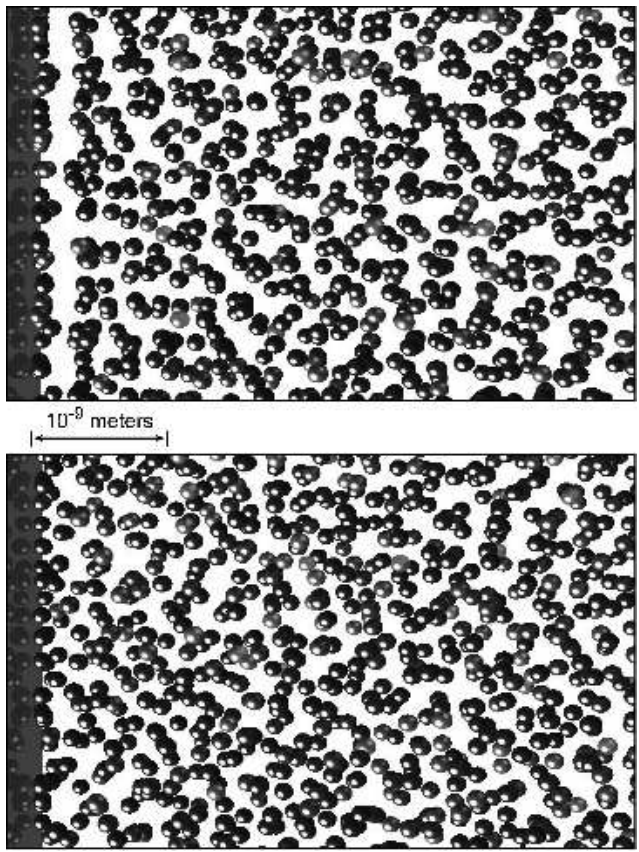,width=3.3in}
\caption{{\bf Piston-induced spatial correlations} -- Two particle slices illustrate the potential problem of spatial correlations developing near the piston boundary. (top) The momentum mirror alone shows a clear particle concentration at the piston surface with a first-neighbor gap clearly developing. (bottom) Adding a region of stochastic forcing ahead of the momentum mirror removes the problem.}
\label{f:piston_bc}
\end{figure}

In our reduced systems, we need to be much more careful with the piston interaction. The piston cannot be allowed to produce spatial correlation, and the purge events must be smooth to ensure that no periodic pulses are introduced to disturb the front dynamics.  Figure \ref{f:piston_bc} demonstrates the problems generated by driving a tin melt with a conventional momentum mirror alone. Driving the same system with a momentum mirror and strong stochastic forcing region is also shown. The spatial correlations are clearly visible without the stochastic forcing (top) and disappear when it is introduced (bottom). Weak Langevin thermostats showed results similar to those of the momentum mirror alone.  We found that damping coefficients very nearly equal to 1/timestep were necessary to prevent the effect.  For these studies the Langevin thermostat was implemented with particle damping and random kicking.  The amplitude of the kicking was set by the system parameters, using the modified equation of motion, 

\begin{equation}
{\bf F} = m{\bf a} - mb{\bf v} + \xi{\bf r}
\end{equation}
\noindent where
\begin{equation}
\xi = \sqrt{\frac{6mbk_BT}{\Delta t}}
\end{equation}
is the amplitude of the stochastic force and ${\bf r}$ is a randomized dimensionless vector with components ranging from -1 to 1.  Here $m$ is the mass, $b$ is the damping coefficient, $k_B$ is the Boltzmann constant, $T$ is the temperature and $\Delta t$ is the timestep.
 
It should be emphasized that the purge preserves the statistical properties of the unpurged boundary, but will not exactly reproduce particle trajectories. Our goal is, first, to accurately reproduce the velocity distributions at the purge point; and second (and more applicable here), to produce as small a disturbance as possible at the purge event.  The momentum mirror with strong stochastic forcing accomplishes both goals.

\vspace{0.25in}
\noindent {\bf Ramping Shock Strength}\\
The final element of the method is to introduce a quasistatic change in the shock forcing parameter.  The goal is to alter the shock strength, while always maintaining a direct mechanical coupling between the forcing at the piston and the response at the shock interface.  The piston velocity increases or decreases by a set amount per timestep.  The point of forcing remains at appoximately a constant distance behind the shock front.  This process continues until the shock strength parameter has reached a desired value, at which point the simulation terminates.

Finally, a second conventional reference run is compared with the final state of the shock ramp.  This allows the final state obtained from the Continuous Hugoniot Method to be compared directly to a conventional shock run to identify any problems and to serve as an error check.  In our experience, the results of the final benchmark runs have been indistinguishable from those obtained from the Continuous Hugoniot Method.

There are two conditions required for the validity of our method: first, that the shock must be steady; and second, that the state at the purge point must be in equilibrium.  The second is guaranteed by our thermalization test before we reduce the system.  The first is guaranteed, so long as our quasistatic ramp loading is slow enough.  If the ramp rate is too fast, the system is not steady.  However, If the ramp rate is quasistatic, small changes have time to equilibrate across the entire system, and thus the driving and response are mechanically coupled.  If this coupling is not maintained, the foundation of the Hugoniot-Rankine equations is eroded, and off-Hugoniot states are produced.  Our ability to very slowly ramp the shock strength parameter while simultaneously assuring that the driving is mechanically coupled to the shock response is an essential property of this method.  We accomplish it by driving our system from a set distance behind the front.  Note that there is only a superficial similarity between our method and isentropic loading experiements where the driving is applied at an ever-increasing distance.  If one ramps the shock driving velocity in an experimental situation, the result is isentropic compression rather than a shock response\citep{rosenberg.01}, as in our technique.

We can estimate an upper bound for the rate at which the shock strength control parameter can be varied.  We nondimensionalize the velocity by the longitudinal sound wave speed behind the shock front $C_S$, and measure time in units of the time needed for sound waves to travel from the rear of the system  to the front and back again:

\begin{equation}
\tilde{v}_p = \frac{v_p}{C_S} \quad \quad \mathrm{and} \quad \quad \tilde{t} = \frac{t}{2L/C_S}
\end{equation}

\noindent where $C_S$ is the sound speed in the shocked state behind the front.  The condition for a quasistatic ramp is then given by
\begin{equation}
\dot{\tilde{v}}_p = \frac{d \tilde{v}_p}{d \tilde{t}} =  \frac{d (v_p/C_S)}{d (t/\frac{2L}{C_S})} \ll 1
\end{equation}
and thus gives the condition on the velocity ramp rate:
\begin{equation}
\frac{dv_p}{dt} \ll \frac{C_S^2}{2L} \quad \Rightarrow \quad \frac{dv_p}{dt} \ll \frac{C_o^2}{2L}
\label{e:ramp_rate}
\end{equation}
where the final step substitutes $C_o$, the ambient uncompressed wave speed, for $C_S$ since it is a more easily calculated quantity and further bounds the value of the velocity ramp rate.

Note that the maximum allowed rate that maintains mechanical coupling between the forcing and response goes to zero for large systems.  Thus as $L$ increases one not only must deal with increasing numbers of particles, but one must also carry out increasingly longer runs.

\section{Application to Lennard-Jonesium}

As a first application, we employ the Lennard-Jones potential, which has been widely used to study shock waves in solids and liquids \cite{holian.88,paskin.77,holian.95,belonoshko.97,holian.98,holian.79,lennard-jones.32}.  Although simple in form and application, the Lennard-Jones potential exhibits much of the phenomenological shock complexity of more ``realistic'' potentials.  We selected it for two primary reasons. First, there is an extensive literature on its shock response, including a published Hugoniot and Hugoniot fit in $U_s$--$v_p$ space; second, the potential is very fast to compute.

\vspace{0.25in}
\noindent{\bf LJ Simulation Details} --
We chose to use the cubic-spline Lennard-Jones 6\,--\,12 potential of Holian \cite{holian.91} in order to allow easy comparison with published Hugoniot results of Germann et al. \cite{germann.00}.  The spline method uses the conventional Lennard-Jones 6\,--\,12 form 
\begin{equation}
\phi_1 = 4 \epsilon \left[ \left(\frac{\sigma}{r} \right)^{12} - \left(\frac{\sigma}{r} \right)^{6} \right]
\end{equation}
for radii from zero to the inflection point, $r_\mathrm{spl}$, which is just beyond the maximum well depth.  Between $r_\mathrm{spl}$ and a point $r_\mathrm{max}$, the potential is given by 
\begin{equation}
\phi_2 = - a_2(r_{\mathrm{max}}^2 - r^2)^2 + a_3(r_\mathrm{max}^2 - r^2)^3
\end{equation}

\noindent The conditions for smoothness of the energy and force at the inflection point uniquely determine $a_2=0.5424494\sigma$, $a_3=0.09350527\sigma$ and $r_\mathrm{spl}=1.244455\sigma$, $r_\mathrm{max}=1.711238 \sigma$, where $\sigma=3.405~\mathrm{\AA}$.

The shock was oriented along the $\langle100\rangle$ direction of the face-centered-cubic crystal structure with unit cell dimension $5.314\,\,\mathrm{\AA}  = 1.561 \sigma$.  Initial temperature was varied with a weak Langevin thermostat from zero to $10 \mathrm{K} = 0.083\,\,k_B\mathrm{T}/\epsilon$.  Results are found not to depend on the initial temperature for shock driving velocities, $v_p$ above $0.75\,C_o$.  Systems were 20 $\times$ 20 lattice planes in cross-section with transverse periodic boundary conditions.  The integration timestep was 0.3 femtoseconds.

Conventional shock simulation runs (benchmark runs) were driven by a warm impactor method and reached $600\,\,\mathrm{\AA}$ in length, consisting of approximately 100,000 particles.  Continuous Hugoniot Method runs were held to $200\,\,\mathrm{\AA}$ in length, consisting of approximately 20,000 particles at any one time.  The effective system size of these treadmilling runs was almost 1.3 $\mu$m in length, consisting of approximately 1,280,000 particles.  In all cases the shocks were given time to establish a steady state (usually 30 to 60 ps).

In this section, we concentrate our efforts primarily in the strong shock (over-driven) regime, $v_p \ge 0.75\,C_o$.  This single-shock regime presents the fewest complications to the co-moving window technique we have developed.  Application to double shocks in the elastic-plastic regime is discussed in Section V.

\begin{figure}[htb]
\epsfig{file=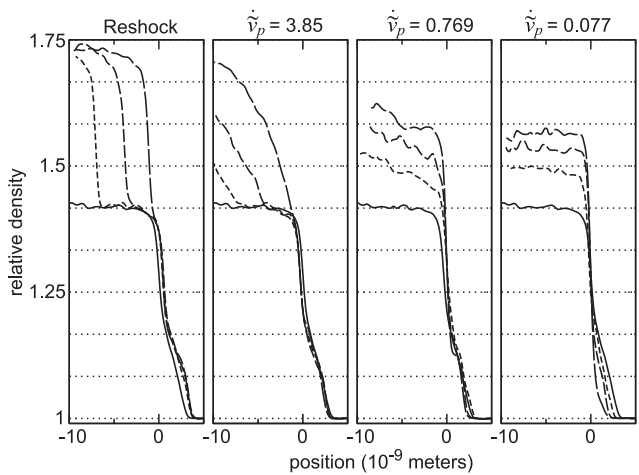,width=3.3in}
\caption{{\bf Ramp rate determination for Lennard-Jones} -- Density profiles for four shock strength ramp rates, ranging from reshock to $\dot{\tilde v}_p=0.077$, are shown.  For each rate, the density profile evolution is shown solid ($v_p=0.75\,C_o$), then three later profiles are shown dashed (short dash $v_p=0.97\,C_o$, medium dashed $v_p=1.19\,C_o$ and long dashed $v_p=1.41\,C_o$).}
\label{f:LJ_ramp_rate}
\end{figure}

\vspace{0.25in}
\noindent {\bf Piston Velocity Ramp Rate Response} --
From Section II, the upper bound for quasistatic ramp rate is
\begin{eqnarray}
{\dot{v}}_p \ll \frac{C_o^2}{2L} & = & 4.5 \times 10^{13}\,\mathrm{m/s^2} \\
                             & = & 0.013\,\mathrm{m/s \,\, per \,\, step}
\end{eqnarray}
\noindent where the wave speed is given by $C_o=\sqrt{72 \epsilon/m}$ \cite{rapaport.95}, $\epsilon=119.8\ k_B$ T and $m=6.63\times 10^{-26}$kg.  $L=200~\mathrm{\AA}$ is the length of the system in these runs.  Figure \ref{f:LJ_ramp_rate} shows the results for four ramp rates, on a piston velocity ramp from $v_p=0.75\,C_o$ to $1.5\,C_o$.  The test ramp rates range from ${\dot{\tilde v}}_p=\infty$~(reshock) to 0.077.

The quasistatic regime is indicated by density profiles which are flat and steady.  The first three profiles are not in the quasistatic regime.  The fourth, with non-dimensional ramp rate ${\dot{\tilde v}}_p= 0.077$ (${\dot{v}}_p=0.001\,\,\mathrm{m/s}$ per step), indicates quasistatic loading for this material and system size.

The ramp rate of $0.001\,\mathrm{m/s \,\, per \,\, step} = 3.3 \times 10^{12}\,\mathrm{m/s^2}$ (corresponding to ${\dot{\tilde v}}_p= 0.077$) will be used for the remainder of the Lennard-Jones work presented.

\subsection{Principal Hugoniot Results}
\begin{figure}[htb]
\epsfig{file=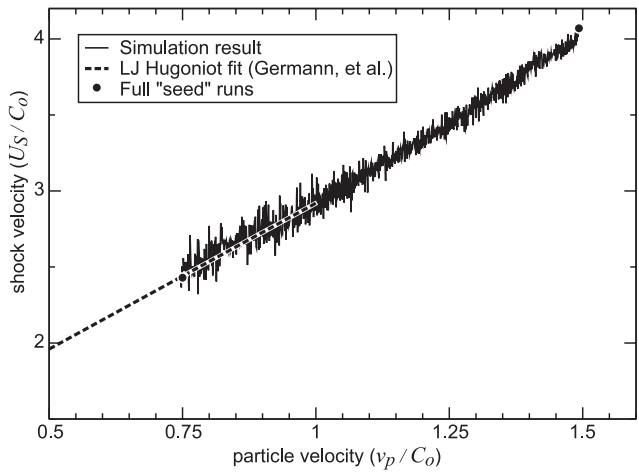,width=3.3in}
\caption{{\bf Continuous principal Hugoniot for Lennard-Jones} -- The results of an application of the Continuous Hugoniot Method to a Lennard-Jones system in the range from $v_p=0.75\,Co$ to $v_p=1.5\,Co$ is shown.  A plot of the published Hugoniot fit of Germann et al. \cite{germann.00} is plotted in its applicable range.  The initial and terminal reference runs are also plotted.  Very good agreement is found between the results of our method and those of published conventional simulations.}
\label{f:LJ_hugoniot_ramp}
\end{figure}

The Continuous Hugoniot Method allows a system to move directly from shock state to shock state.  Therefore, the path of our Continuous Hugoniot Method through $U_s$--$v_p$ (or $P$--$V$) space during a single run is the principal Hugoniot of final shock states in the material.  This is true to the extent that the system's values within its reduced system are indicative of the values very far behind the shock (i.e. there is convergence to the final shock state within a short distance behind the shock front).  Figure \ref{f:LJ_hugoniot_ramp} shows such a path for a run in Lennard-Jones, beginning at $v_p=0.75\,C_o$ and running continuously through $v_p=1.5\,C_o$.  Initial and terminal reference runs, made independently at $v_p=0.75\,C_o$ and $v_p=1.5\,C_o$, respectively, are also shown.  Finally, the Hugoniot fit proposed by Germann et al. \cite{germann.00} is plotted over its applicable range.

We see very good agreement of our simulation results with both comparisons.  In the lower range of $v_p$ we can compare to the published fit.  We see here that our simulation data overlays the fit very nicely and continues the line of the fit beyond the range for which it was originally published.  At higher values of the piston driving velocity we see our simulation data stiffens as it should showing a super-linear increase in $U_S$ vs $u_p$.  In the upper range of $v_p$ there is no published fit, but we can compare to our reference runs.  We can see that the Continuous Hugoniot Method data and the terminal reference results agree acceptably.

We would expect that if the Continuous Hugoniot Method were following some other path than the Hugoniot, that this would become more apparent at later points in the simulation runs (i.e. the error would accumulate).  If for instance, we were following a thermodynamic path such as an isentrope the path would diverge and fall below the Hugoniot later in the simulation run.  This divergence from the Hugoniot states would therefore be most obvious late in our simulations.  We do not see this effect to any significant degree in our data.  Moreover, we see excellent agreement between the system state of the ramped system and the steady reference state at the terminus.

\subsection{Comparison of Density Profiles}
\begin{figure}
\psfig{file=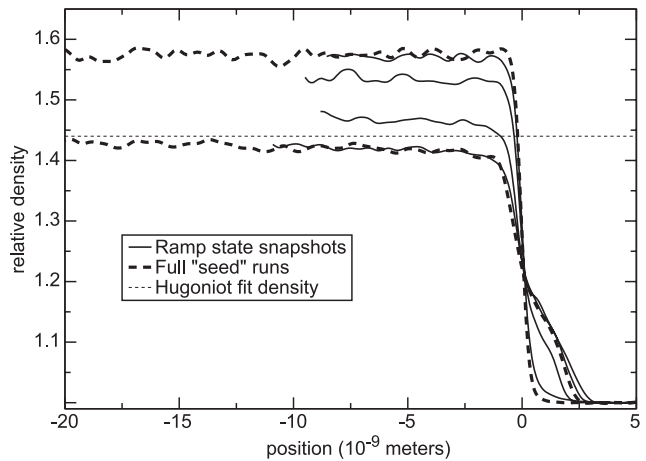,width=3.3in}
\caption{{\bf Density comparisons for Lennard-Jones} -- This plot shows density profile snapshots taken from a shock strength ramp (black solid) and compares them with density profiles from the initial and terminal seed runs (thick dashed) and those predicted by the Hugoniot fit of Germann et al. (thin dashed).  We see near perfect agreement between our ramp data and our reference runs.  The ramp data is, however, consistently below the densities predicted by Germann.  We attribute this to slow spatial convergence of the density.}
\label{f:LJ_profile_comparison}
\end{figure}

Figure \ref{f:LJ_profile_comparison} shows a series of five density profile snapshots taken from a single ramp run (shown as solid lines).  They are equally spaced in the piston velocity from $0.75\,\,C_o$ to $1.5\,\,C_o$. Also plotted are the density profiles of the two reference runs.  Using the fit equation and the mass conservation jump condition, one can derive the density for the $0.75\,\,C_o$ state.

\begin{equation}
\frac{\rho}{\rho_o} = \frac{U_s}{U_s-v_p} = \frac{1+1.92v_p}{1+0.92v_p}
\end{equation}

For $v_p=0.75\,\,C_o$, the predicted relative density $\rho/\rho_0$ of 1.44 is plotted as a thin dashed line in Figure \ref{f:LJ_profile_comparison}.  The remaining profiles fall above the applicable range of the published fit.

We see that the density predicted by the published fit is something less than 2\% above the average density of the profiles from the Continuous Hugoniot Method.  However, in the case of the $0.75\,\,C_o$ density profile, we see ideal agreement with the seed profile up to $100\,\,\mathrm{\AA}$ behind the front.  Beyond $100\,\,\mathrm{\AA}$, where the results of our method have been purged, the seed density continues to approach the fit prediction.  We see the same results in a comparison with the $v_p=1.5\,\,C_o$ seed run.  This observation leads us to the conclusion that our method is accurately predicting the Hugoniot state density profile within the spatial extent of the reduced system.  However, densities do not necessarily converge within the reduced systems to the values predicted far behind the shock front.  Therefore, comparison between Hugoniot density and the average density within our reduced system cannot be made directly.  If comparison is desired, the density profile of the reduced system would need to be extrapolated, or the reduced systems would need to be enlarged.

\subsection{Comparison of Final States}
As previously noted, the final state of the system is a particularly important point of comparison because it is the state of the maximum integrated error.  Direct comparison of every state of the system would be impractical due to the computational cost of benchmark runs.  We can, however, cautiously assume that good agreement of the final state indicates good agreement throughout.

\begin{figure}[htb]
\epsfig{file=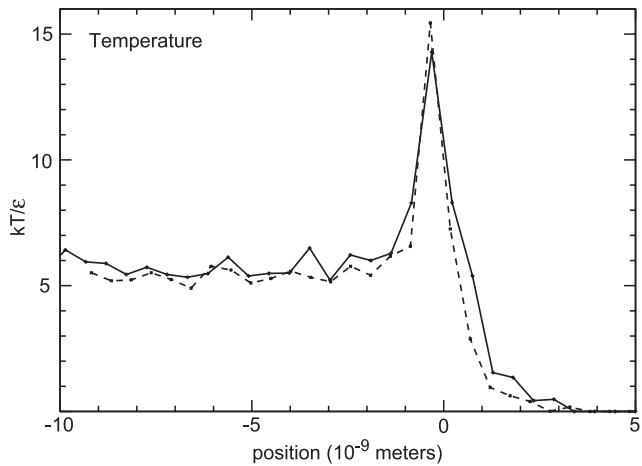,width=3.3in}
\caption{{\bf Temperature comparison} -- The temperature profiles for states obtained from the Continuous Hugoniot Method (dashed line) and the terminal seed run (solid line) are plotted for shock strength $v_p = 2000\,\,\mathrm{m/s} = 1.5\,\,C_o$ in a Lennard-Jones system.  We see very good agreement in the peak temperature and final temperature behind the front.}
\label{f:LJ_thermo_comp_2000}
\end{figure}

\begin{figure}[htb]
\epsfig{file=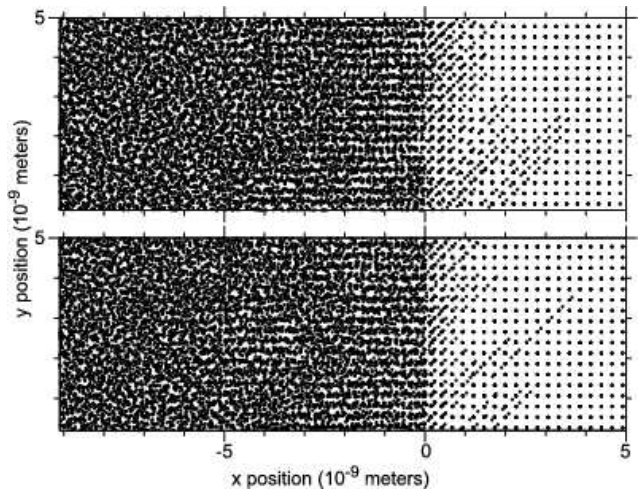,width=3.3in}
\caption{{\bf Structural dynamics comparison} -- Particle slices in the region of the shock front for states obtained from (bottom) the Continuous Hugoniot Method and (top) the terminal seed run  at shock strength, $v_p = 2000\,\,\mathrm{m/s} = 1.5\,\,C_o$ in a Lennard-Jones system.   We see very good qualitative agreement in the deformation behind the front with pockets of order mixed with disorder.  We also see the development of forward reaching structures ahead of the shock front.}
\label{f:LJ_particle_comp_2000}
\end{figure}

\begin{figure}[htb]
\epsfig{file=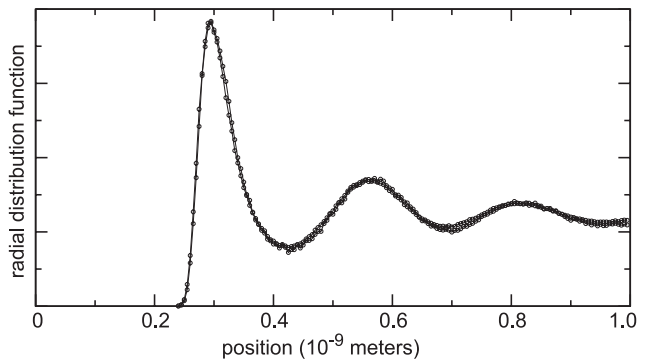,width=3.3in}
\caption{{\bf Radial distribution function comparison} -- RDF for the compressed material extending 10 nm behind each front is shown for the Continuous Hugoniot Method (dashed line) and the terminal seed run in (solid line) at shock strength, $v_p = 2000\,\,\mathrm{m/s} = 1.5\,\,C_o$ in a Lennard-Jones system.}
\label{f:LJ_rdf}
\end{figure}

We have already seen that the Continuous Hugoniot Method produces density profiles which concur with conventional reference runs.  Figure \ref{f:LJ_thermo_comp_2000} shows a comparison of the temperature profiles for the final system states.  Here, the results of the Continuous Hugoniot Method are shown dashed and the results of the reference run are shown solid, both for a piston velocity $v_p=2000\,\,\mathrm{m/s}=1.5\,C_o$.  The zero position coincides with the shock front location.  We note, that the temperature peaks are in good agreement and the temperature relaxes similarly to identical values within the system fluctuations.

Figure \ref{f:LJ_particle_comp_2000} illustrates particle snapshots from the terminal reference state (top) and the Continuous Hugoniot Method (bottom).  The slices are along the $x$--$z$ plane at the final piston velocity $v_p=2000\,\,\mathrm{m/s}=1.5\,C_o$.  The particle snapshots clearly illustrate a qualitative agreement between the two systems' dynamic structures.  We see that both systems exhibit a strong disordering transition on similar length scales, similar islands of incomplete disordering, and similarly sharp density transition.  Both have also developed forward-reaching features ahead of the front.  In Figure \ref{f:LJ_rdf}, we see the radial distribution function for the material behind the front.  The rdf results for the Continuous Hugoniot Method are shown dashed, and the results of the seed run are shown in solid.  We see from Figure \ref{f:LJ_particle_comp_2000} both very good qualitative and quantitative agreement of our method with conventional shock methods.

\section{Application to Tin}

This section describes the application of the Continuous Hugoniot Method to a more realistic material potential with the goal of comparing directly with experiments on tin.  We take advantage of the method's increased efficiency in producing a dense sampling of Hugoniot shock states to study processes at the shock front.  

Tin is a natural choice for laser ablation experiments in shock melting because it undergoes shock transitions at relatively moderate laser intensities.  Experiment places its melt on release at $\sim250$ kbar and shock melting at $\sim500$ kbar \cite{mabire.00}.  These pressures are well within the range of terawatt class laser systems.  Moreover, tin's crystal structure makes reflectivity and harmonic diagnostics possible on breakout.

\vspace{0.25in}
\noindent{\bf Modeling Tin with MEAM} --
We selected the Modified Embedded Atom Method (MEAM) as our interatomic interaction model for tin.  This potential has been demonstrated to handle shock simulations \cite{cherne.04}, and has been applied successfully to the study of melt properties of tin \cite{ravelo.97}.

MEAM was first proposed by Baskes in 1992 \cite{baskes.92}, as an extension of the highly successful Embedded Atom Method (EAM) of Daw and Baskes \cite{daw.84}.  Both methods, inspired by density functional theory, compute energies using a semi-empirical combination of two-body interaction and environment-specific electron density embedding energies.  MEAM extends EAM to handle covalent bonding by introducing angle-dependent electron densities.

To visualize the embedded atom, recall the chemist's notion of the electron density clouds surrounding an atom.  These lobe-shaped orbital structures overlap as atoms approach each other and form a background electron density within which each is embedded.  With only two atoms, this energy of embedding within the other's electron density can be absorbed easily into a two-body interaction (i.e. bond).  As atom density increases, however, the electron density becomes a function of many atoms and the embedding energy therefore becomes an effective environment-dependent interaction.

The environment-dependent nature of MEAM makes it especially good for applications near surfaces, voids, defects and interfaces.  In these regions, where the local environment is very different from the bulk environment, many other potentials are at their weakest, having been parameterized with bulk response measurements.  MEAM adjusts well to these situations, being long-range in open structures and short-range in dense closed structures, based on screening.

MEAM has been modified, updated, generalized and expanded with subsequent publications \cite{baskes.92,baskes.94,ravelo.97}. Because the method varies subtly between publications, we specify our implementation in detail in Appendix A.

\vspace{0.25in}
\noindent{\bf Tin Simulation Details} --
The initial state for our runs was a zero-temperature $\beta\,-\,$tin (white tin) in a tetrahedral crystal structure with unit cell dimensions $5.92~\mathrm{\AA}~\times~5.92~\mathrm{\AA}~\times~3.232~\mathrm{\AA}$ \cite{wyckoff.24}.  The shock was oriented to run along the $<$100$>$ direction.  All runs were $10~\times~10$ lattice planes in cross section.  We employed periodic boundary conditions in each of the transverse directions.  The integration timestep was 0.3 femtoseconds.

Most runs were made using the Continuous Hugoniot Method.  These runs were held to a reduced system size of 165 $\mathrm{\AA}$ in length.  Approximately 4,000 particles were simulated at any given time.  Over the course of the entire series of runs, the system had an effective length of 0.26 $\mu$m consisting of more than 178,000 particles participating.  The velocity ramp runs analyzed in this section went from $v_p=2000~\mathrm{m/s}$ to $2300~\mathrm{m/s}$ over the equivalent of 45 picoseconds (150,000 timesteps).

Seed runs, using traditional shock simulation methods, were driven by a warm impactor.  The seed simulations were allowed to run for 9 to 12 picoseconds (30,000 to 40,000 timesteps), until they reached a steady state.  Seed simulation sizes grew to approximately 580~$\mathrm{\AA}$ in length and consisted of approximately 15,000 particles at any one time.

Efficiency of these methods will be addressed in Section VI.
\subsection{Tin Shock Melting and Hugoniot}

\begin{figure}[htb]
\begin{center}
\psfig{file=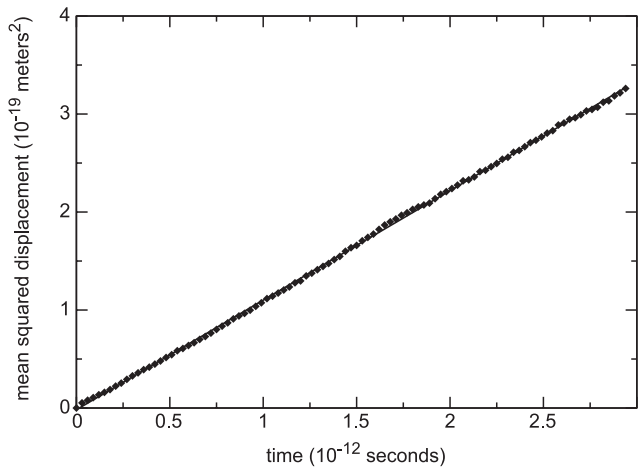,width=3.3in}
\caption[Mean squared displacement indicating tin melt]{{\bf Mean squared displacement indicating tin melt} -- The mean squared displacement (msd) is plotted as a function of time with a linear fit for a 45 $\mathrm{\AA}$ wide slice of material from behind the front for shock strength, $v_p=3000~\mathrm{m/s}$.  The linear increase of the msd with time is a classic indicator of liquid dynamics.}
\label{f:msd}
\end{center}
\end{figure}

Tin undergoes a shock melt and completely thermalizes in simulation on short length scales at moderate shock strengths.  To verify the properties of the tin melt, mean square displacement measurements were made to study the temporal dynamics of the melt.  This is a very good way to distinguish a liquid from a glass or super-cooled state.  Figure \ref{f:msd} shows the results of our analysis for the same $45~\mathrm{\AA}$ thick sample as describe above.

The plot indicates that the material exhibits liquid-like diffusional properties.  Theory predicts that the mean square displacement grows linearly with time in a liquid
\begin{equation}
\left<\left({\bf r}(t) - {\bf r}(0)\right)^2\right> = 6\, D\, t
\end{equation}
where $D$ is the diffusion coefficient.  Here $D=1.84~\times~10^{-8}~\mathrm{m^2/s}$.  Experiments by Cahoon \cite{cahoon.03} found a value of $D=2.58~\times~10^{-8}~\mathrm{m^2/s}$ for standard pressure at similar temperatures using an Arrhenius equation approach to fit experimental tin diffussion data.  High-pressure experiments have not been conducted in molten tin.

\begin{figure}[htb]
\begin{center}
\psfig{file=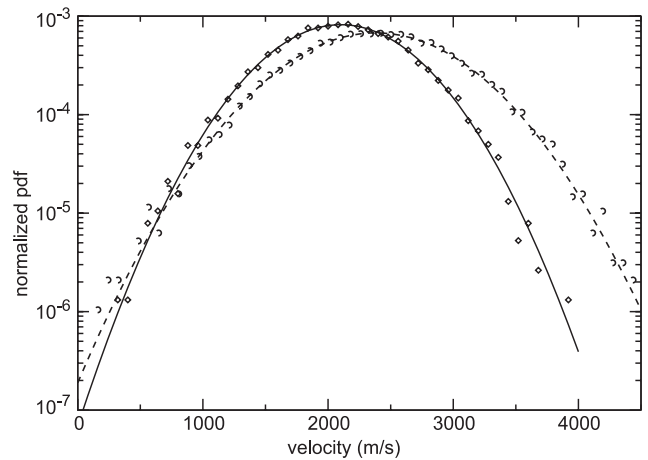,width=3.3in}
\caption[Thermalized velocity distribution functions for tin]{{\bf Thermalized velocity distribution functions for tin} -- Velocity distribution functions in $v_x$ for two approximately 100 $\mathrm\AA$ slices of material approximately from 50 $\mathrm\AA$ behind the front.  Data is shown for $v_p=2000\,\,\mathrm{m/s}$ (solid line) and $v_p=2300\,\,\mathrm{m/s}$ (dashed line) from a series of runs using the Continuous Hugoniot Method in tin.  Gaussian fits are also shown.  The data, plotted in log-linear to accentuate the tails, fits nicely, indicating that the material has thermalized to a Maxwellian distribution.}
\label{f:two_velocity_distributions}
\end{center}
\end{figure}

The velocity distribution functions for the $x$-component of velocity for two moderate shock runs are shown in Figure \ref{f:two_velocity_distributions}.  Data is shown for $v_p=2000\,\,\mathrm{m/s}$ (solid line) and $v_p=2300\,\,\mathrm{m/s}$ (dashed line).  Gaussian fits are also shown.  These two shock states were chosen, as they are the initial and terminal states of a series of Continuous Hugoniot Method runs in tin.  One should notice first that both systems behind the front are well described by equilibrium statistics, even in the tails.  The material has thermalized, a condition necessary for rigorous application of the Continuous Hugoniot Method.

The fit has the form of a Maxwellian velocity distribution \cite{reichl.98},
\begin{equation}
f(v)=A e^{-\frac{m\left(v_x - v_p\right)^2}{2k_bT}}
\end{equation}
where $A$ is a normalization constant, $m=118.7~\mathrm{amu}$ is the particle mass of Sn, and $k_b$ is the Boltzmann constant and $T$ is the temperature.

Results in diamonds are for a piston driving velocity of $v_p=2000\,\,\mathrm{m/s}$.  The fit parameters give values of $2021~\mathrm{m/s}$ and 3360 K, for $v_p$ and $T$, respectively.  Results in semi-circles are for a piston driving velocity of $v_p=2300\,\,\mathrm{m/s}$.  The fit parameters there are $2301~\mathrm{m/s}$ and 4958 K.  These are very good fits considering the relatively small number of particles available for the calculation.  Time averaging over 500 timesteps was used to produce these distributions.

Recall that we can set an upper limit on the piston velocity ramp rate from Equation \ref{e:ramp_rate}.  For tin this gives,
\begin{eqnarray}
\dot{v}_p \ll \frac{C_o^2}{2L} & = & 2.38~\times~10^{14}~\mathrm{m/s^2} \\
                               & = & 0.0714~\mathrm{m/s~per~step}
\end{eqnarray}
where we have used $C_0 \simeq 2800~\mathrm{m/s}$ for the wave speed and $L=165~\mathrm{\AA}$ for the length of the system.  This bound for the quasistatic piston acceleration is more than five times what we derived for Lennard-Jones because the wave speeds in tin are higher and the system sizes are smaller.  Combining this result with the ramp rate testing which we conducted in Lennard-Jones, we determine that piston velocity ramp rates as high as $0.005~\mathrm{m/s~per~step}$ should be allowed in tin without jeopardizing the quasistatic condition of the Continuous Hugoniot Method.

For an added measure of assurance, we use the smaller velocity ramp rate of $\dot{v}_p=0.002~\mathrm{m/s~per~step}=6.67~\times~10^{12}~\mathrm{m/s^2}$ in all of the simulations discussed in this section.

\begin{figure}[htb]
\begin{center}
\psfig{file=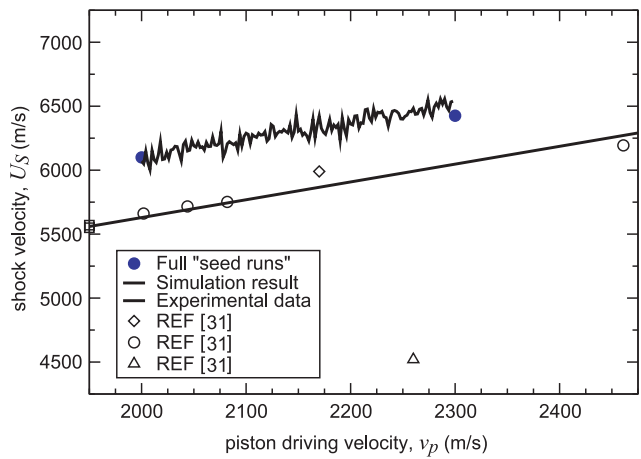,width=3.3in}
\caption[Continuous principal Hugoniot for tin]{{\bf  Continuous principal Hugoniot for tin} -- The results of an application of the Continuous Hugoniot Method to a tin system in the range from $v_p=2000~\mathrm{m/s}$ to $v_p=2300~\mathrm{m/s}$ is shown.  Experimental data are plotted from several sources and a fit is made to the experimental data.  Agreement is to within 6\% across the board, with good agreement in slope.}
\label{f:tin_hugoniot}
\end{center}
\end{figure}

Application of the Continuous Hugoniot Method to tin gives the path through $U_S-v_p$ space shown in Figure \ref{f:tin_hugoniot}.  These results are compiled from a series of runs beginning at $v_p=2000~\mathrm{m/s}$ and running continuously through $v_p=2300~\mathrm{m/s}$.  Initial and terminal reference runs, made independently, are also plotted.  Experimental results from the literature are plotted and a linear fit is given to the experimental data.  The initial shock strength parameter of $v_p=2000~\mathrm{m/s}$ was selected to assure that the results were within the strong shock regime.

For experimental comparison, $v_p=2000~\mathrm{m/s}$ gives a pressure of $730~\mathrm{kbar}$ and $v_p=2300~\mathrm{m/s}$ gives a pressure of $850~\mathrm{kbar}$.

We can see that the $U_S$ versus $v_p$ relationship is approximately linear in this range, as in Lennard-Jones.  The $U_S$ response is noisy, but not significantly more noisy than traditional shock simulation.  Although experimental data is not currently available for single-crystal tin, we were able to compare to room-temperature polycrystalline tin shock experiments.  One can see relatively good agreement with experimental data from the literature \cite{marsh.80, russianweb}, however our results are consistently above experimental values.  Agreement with experimental data is to within 6\% across the board, and shows good agreement with the slope.

Given that Ravelo and Baskes found the MEAM potential to match transition temperatures to within 11\%, our measurements are likely the best agreement that the potential can provide.  These results are rather encouraging to further comparison between our computation results and experiment.

\subsection{Continuous Temperature Profiles versus Shock Strength}

Our method allows study of the shock front at a fine resolution in the shock strength parameter.  Where traditional simulation and experiment provide discrete Hugoniot points, our method provides a continuum of states.  This may be particularly useful in the study of melting and other continuous transitions.

\begin{figure}[htb]
\begin{center}
\psfig{file=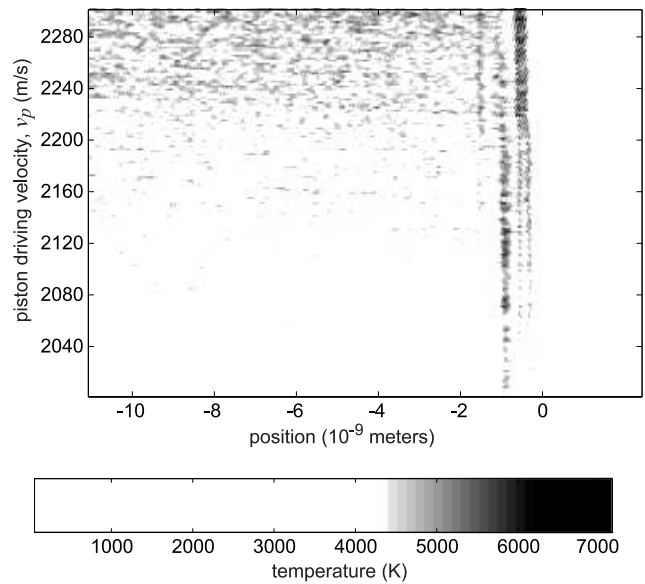,width=3.3in}
\caption[Temperature profile versus shock strength for tin]{{\bf  Temperature profile versus shock strength for tin} -- Temperature is represented as a function of both position and shock strength.  Temperature is represented as a third dimension using gray scale.  The reference frame is moving with the shock front, so that the front is always at $x=0$.  The shock is moving to the right with cold (white) pristine material ahead of it.}
\label{f:temp_shock}
\end{center}
\end{figure}

We apply the method first to characterize the temperature profiles in tin as a function of shock driving velocity.  Figure \ref{f:temp_shock} shows temperature as a function of position in horizontal bands, with the shock strength increasing up the plot.  The shock is moving left to right.  The white block on the right is the low temperature initial state.  Temperature profiles were spatially averaged as described by Hardy \cite{hardy.82}.

\begin{figure}[htb]
\begin{center}
\psfig{file=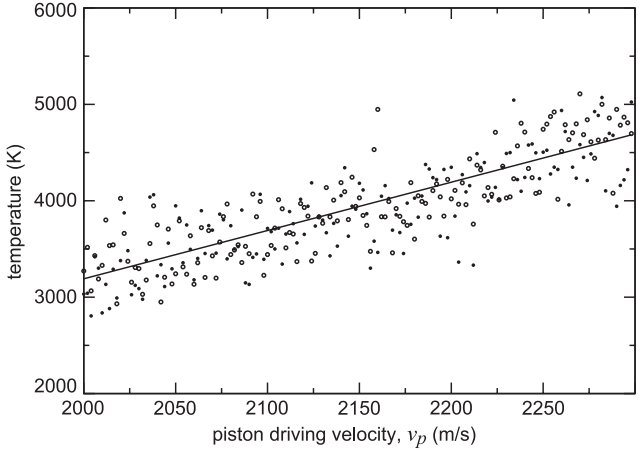,width=3.3in}
\caption[Final Temperature versus shock strength for tin]{{\bf Final Temperature versus shock strength for tin} -- Temperature is plotted as a function of shock strength, $v_p$, at a distance of $50~\mathrm{\AA}$ behind the shock front (hollow).  A linear fit is provided.  Also plotted are temperatures taken at constant time ($t=.83~\mathrm{ps}$) behind the shock front (solid).}
\label{f:temp_slice}
\end{center}
\end{figure}

Note first that the temperature behind the front increases gradually with shock strength.  Figure \ref{f:temp_slice} shows a slice at a distance $50~\mathrm{\AA}$ behind the front.  We can see that the temperature in this region is noisy, but follows a clear trend, growing linearly with the shock strength in this range.

\begin{figure}[htb]
\begin{center}
\psfig{file=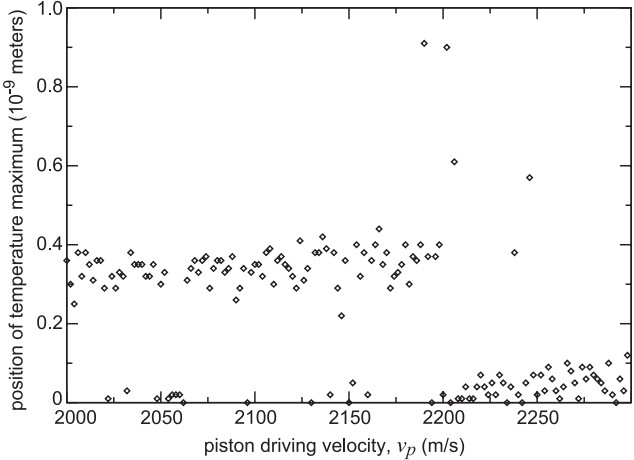,width=3.3in}
\caption[Peak temperature position versus shock strength for tin]{{\bf  Peak temperature position versus shock strength for tin} -- The position, measured from the shock front, of the temperature profile peak is plotted as a function of shock driving velocity, $v_p$.}
\label{f:peak_temp_shock}
\end{center}
\end{figure}

More interesting are the temperature profiles within the shock front and near the temperature peak.  This temperature peak is a common feature of the shock interface.  Our data seem to indicate the peak position is stable for most of the shock strength ramp.  But, at approximately 2200 m/s, the position of the peak shifts forward abruptly.  These results can be seen more directly in Figure \ref{f:peak_temp_shock}.  We do not have an explanation for the behavior.

\subsection{Melt Length Scale}
The primary comparison point between simulation and experiment in this work is the length scale for the melt process behind the shock front.  Experimental work is proposed that could measure the timescales between back surface arrival of the shock and the melt to within a few tens of femtoseconds.  This equates to a spatial resolution of only a few tens of Angstrom.  There is, therefore, great interest to establish a relationship between the melt length scale and the shock strength, which we have parameterized by the piston velocity, $v_p$.

\begin{figure}[htb]
\begin{center}
\psfig{file=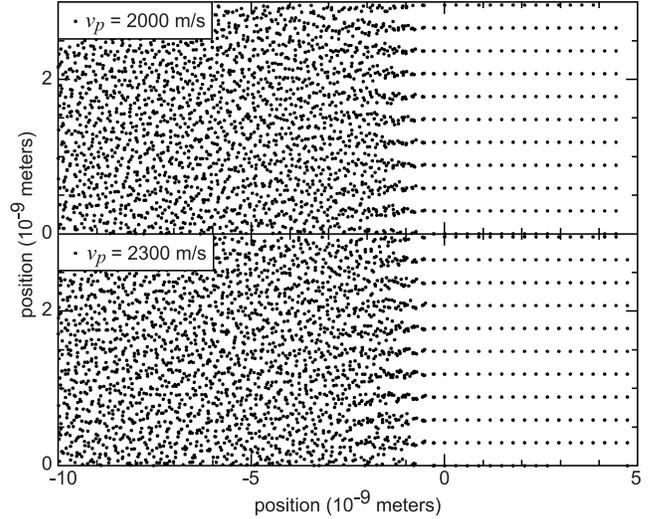,width=3.3in}
\caption[Particle melt dynamics comparison]{{\bf Particle melt dynamics comparison} -- Particle snapshots from the region near the shock front where the melt is occurring.  The length scale for melting can be seen as the lattice planes gradually spread to fill the volume uniformly.  The eye has a difficult time comparing the melt scales.}
\label{f:melt_profiles}
\end{center}
\end{figure}
 
One can see in Figure \ref{f:melt_profiles} particle profiles for the low ($v_p=2000~\mathrm{m/s}$) and high ($v_p=2300~\mathrm{m/s}$) shock strength in our ramp.  On the right in each is undisturbed crystal and on the left is the shock-induced melt.  The transition between the two is notoriously difficult to characterize, because the nature of the melt is nonlocal.  Exactly locating a collective response such as a melt point is nontrivial, especially with noisy systems.

Proper quantitative analysis requires calculation of a mean square displacement as a function of distance from the front.  However, no quantitative trend between melt length scale and shock strength has been able to be derived from the analysis, due to excessive noise to signal within the system.

\begin{figure}[htb]
\begin{center}
\psfig{file=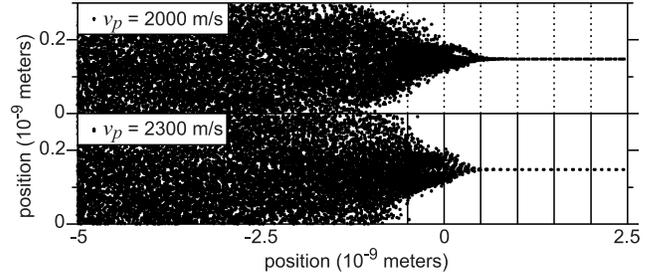,width=3.3in}
\caption[Particle collapse]{{\bf Particle collapse} -- Collapse of all lattice planes onto each other indicates the length scale for the diffusion of particles to fills space.  The Lindemann criterion for the onset of 2D melting is met when the mean free displacement is greater than a given fraction of the crystal spacing.}
\label{f:melt_collapse}
\end{center}
\end{figure}

A more visual analysis demonstrates the problem.  Figure \ref{f:melt_collapse} shows the same data collapsed to a line and averaged over 1000 timesteps.  We can now begin to see the length scale to melt developing.  If we have in mind something like a Lindemann criterion \cite{gilvarry.56} for the onset of disorder, we can argue that the melt has occurred when a particle from one lattice plane has diffused to a degree that it cannot be differentiated from a particle from an adjacent plane.  We thereby define the melt point where the particles have full coverage of the cell in Figure \ref{f:melt_collapse}.  Thus we can see that the melt occurs on a shortened length scale for $v_p=2300~\mathrm{m/s}$ by approximately 3 to $5~\mathrm{\AA}$, compared to $v_p=2000~\mathrm{m/s}$.  It is clear, now, how this signal could be easily swallowed by the noise of the system.  Further, such a length scale could not be resolved experimentally.

We conclude that the melt front in tin is too abrupt to be distinguished from the shock front at shock strengths within the strong shock regime.  We are challenged to move into the weaker elastic-plastic shock regime, where the two fronts diverge with time.

\section{Two-shock states and the Strong to Elastic-plastic Shock Transition}

We begin this section with a brief review of the three distinct regimes of shock propagation in simple solids.  The analysis begins with the shock Hugoniot.  For an excellent older review of experimental measurements and theoretical fits, see Rice, McQueen and Walsh \cite{rice.58}.  They provide examples of experimentally determined Hugoniots, EOS models, and the gamut from descriptions of experimental techniques to basic theory on shock waves in solids.

\begin{figure}[htb]
\begin{center}
\psfig{file=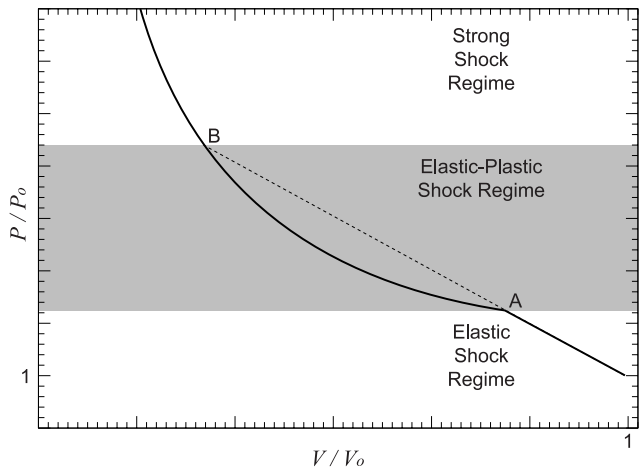,width=3.3in}
\caption[Shock regime phase space]{{\bf Shock regime phase space} -- The three standard shock regimes for a solid are identified.  The point A is the Hugoniot Elastic Limit where the material begins to deform plastically due to shear stresses.  The point B is the transition point on the Hugoniot between the strong shock and elastic-plastic shock regimes.}
\label{f:regimes_hugoniot}
\end{center}
\end{figure}

At high shock strength, the material response is considered over-driven.  In this case, a single plastic/melt wave propagates through the material.  The shock front shows a steep gradient in all thermodynamic properties.

As the shock strength is reduced, the material response changes and the material supports two waves.  There is a primary plastic wave front, which is preceded by a faster elastic precursor shock.  With time, the two waves move apart.  The division of the shock can be explained by looking at the Hugoniot: see Figure \ref{f:regimes_hugoniot}.  Points on the Hugoniot below the point B are reached by moving from the initial condition to point A along a straight line.  Then a second straight line connects point A to the final state on the Hugoniot curve.  One can show from the jump conditions in Section I that the slopes of these two lines are proportional to the velocities of the two wave fronts.  In the case of the strong shock (above point B on the Hugoniot), a single line connects the initial condition and the final Hugoniot state.  Thus there is a single front. 

For stresses below the point A in Figure \ref{f:regimes_hugoniot}, the solid supports completely elastic, reversible shock waves.  In this regime, transient shock waves which compress and release a material leave no permanent distortion.  We do not address the elastic regime in this work, but assert that the Continuous Hugoniot Method could be applied to that regime without difficulty.

\vspace{0.25in}
\noindent{\bf Applicability to the two-wave shock} --
To this point, we have investigated only the strong shock regime.  Here we will extend the technique into the elastic-plastic regime and characterize the transition.  However, it should be made clear that there is some potential difficulty in so doing.

In Section II, we showed that the upper bound for the velocity ramp rate used was inversely proportional to the system size.  This is not a serious hindrance in the strong shock regime because we have shown that the system can be continually purged, given that the shocked material is known to thermalize within a fixed distance behind the front.  In the two-wave regime we have just discussed, however, the elastic precursor can theoretically move arbitrarily far ahead of the plastic/melt wave.  In fact, it can be shown that the precursor front {\it must} have a higher velocity than the plastic front.  Thus the system grows in time, and long runs must be simulated with large systems.

The assumptions of the method never break down, but the computational efficiency advantage of the method quickly deteriorates.  Larger systems give slower ramp rates, which take longer times, which require larger systems.  We can see from this argument that the method must be applied with care to the elastic-plastic regime.

\vspace{0.25in}
\noindent{\bf Results} --
The results presented here are for a decreasing Continuous Hugoniot Method run in tin beginning from a shock state just within the strong shock regime.  The initial state has a shock piston velocity $v_p = 2000~\mathrm{m/s}$ and runs into the elastic-plastic regime to a final state at $v_p = 1250~\mathrm{m/s}$.

\begin{figure}[htb]
\begin{center}
\psfig{file=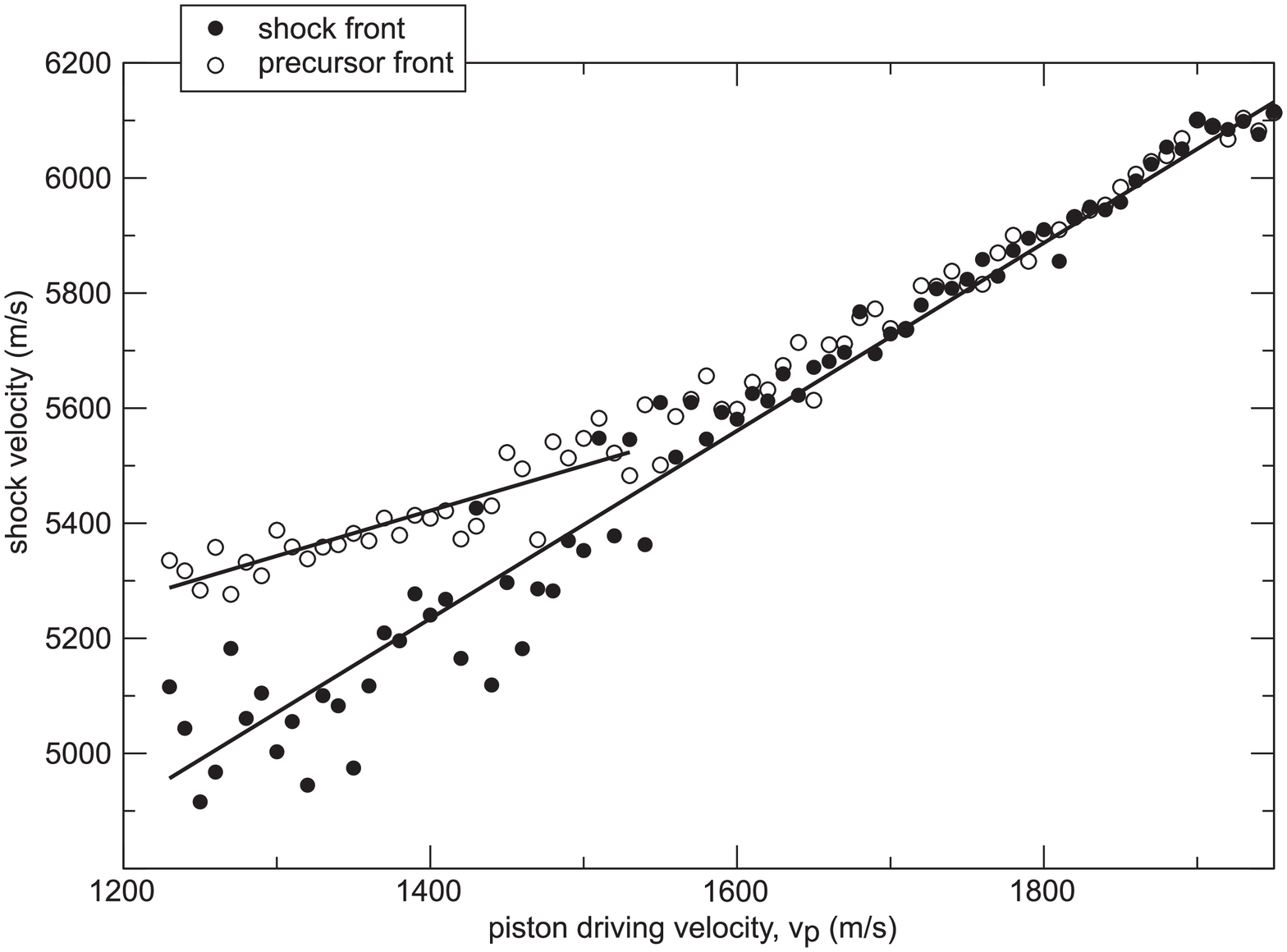,width=3.3in}
\caption[Tin shock and precursor velocities]{{\bf Tin shock and precursor velocities} -- Shock Hugoniot plot for a decreasing Continuous Hugoniot Method run from 2000 m/s to 1250 m/s which shows the emergence of the elastic precursor wave.}
\label{f:hugoniot_split}
\end{center}
\end{figure}

Figure \ref{f:hugoniot_split} shows the Hugoniot representation for the run.  The plot shows that at high shock strength there is no precursor ahead of the shock front, but that an elastic precursor forms at lower forcing.  The transition point comes at approximately 1560 m/s in piston velocity.

We can recast the data by defining an order parameter as the difference between the shock front and elastic precursor velocities.
\begin{equation}
\Phi= U_{S,precursor} - U_{S}
\end{equation}
and the nondimensional bifurcation parameter
\begin{equation}
\epsilon= \frac{u_p}{u_{p,\mathrm{critical}}} - 1
\end{equation}

\begin{figure}[htb]
\begin{center}
\psfig{file=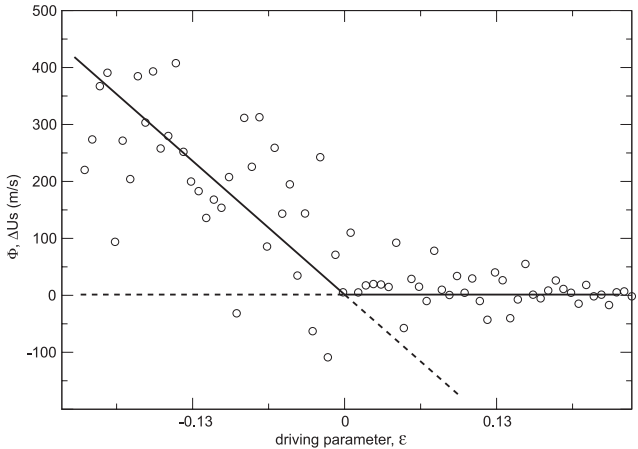,width=3.3in}
\caption[Shock regime bifurcation]{{\bf Shock regime bifurcation} -- Shock regime order parameter, $\Phi = U_{S,precursor} - U_{S}$ vs. the shock control parameter.  The data is consistent with a transcritical bifurcation.}
\label{f:transcritical}
\end{center}
\end{figure}

Figure \ref{f:transcritical} shows the results of the transformation.  Solid and dotted lines are overlayed to represent the stable and unstable branches of a transcritical bifurcation, which seems the most likely conventional scenario to decribe this noisy system.  The bifurcation shows the classic linear increase from zero at onset.  The bifurcation is continuous and imperfect.

\begin{figure}[htb]
\begin{center}
\psfig{file=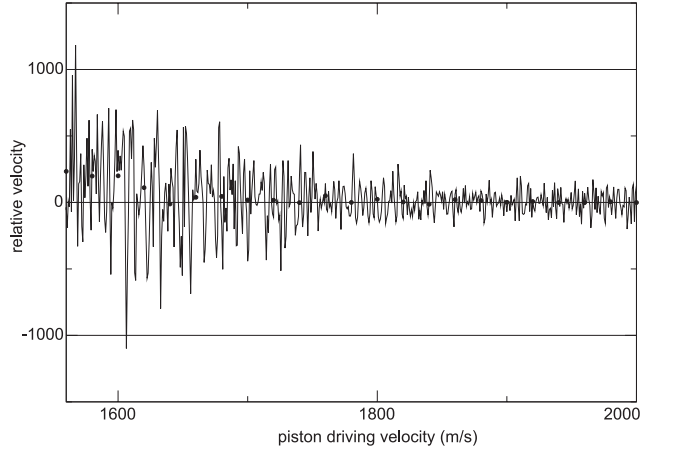,width=3.3in}
\caption[Fluctuation at onset]{{\bf Fluctuation at onset} -- The fluctuation in the order parameter increases as the system approaches onset from above.}
\label{f:fluctuation}
\end{center}
\end{figure}

Analysis of fluctuations in approaching onset from above shows a characteristic increase in fluctuation magnitude.  Figure \ref{f:fluctuation} shows the standard deviation of the order parameter above onset.  The inherently noisy nature of the system makes it possible to fit either a linear or sublinear function to the data.

\section{Efficiency and speed up}
A rigorous comparison of processor timing between conventional methods and our Continuous Hugoniot Method is perhaps not possible, because the nature of the output data is different.  Our method gives a much more finely spaced collection of shock states than conventional methods.

In the Lennard-Jones system, we saw the computation time to compute the two reference runs by conventional methods was approximately equal to the computation time necessary to compute the entire intermediate Hugoniot, via the Continuous Hugoniot Method.  This can be a significant speed-up when a dense sampling of states is necessary.  The impact of the method is also felt in memory and disk resources where we were able to simulate the cumulative effect of 1,280,000 particles with the resources required to hold only 20,000 at any one time.

The complexity of the MEAM potential is substantially greater than that of Lennard-Jones.  This complexity translates into added computation time as processors deal with the additional instructions and the multi-tier cascade of nested loops.  For moderate simulation sizes such as $10~\times~10~\times~40$ lattice planes, MEAM has been clocked as running between 90 and 200 times slower than a comparable system of Lennard-Jones interactions.  The factor is significant since it increases shock run computational wall clock times from days to months or years.  The payoff is, of course, that MEAM is a relatively accurate model for a real material.  As such, results can be compared to experiment.

Without the Continuous Hugoniot Method, progress in tin might have been prohibitively slow.  The data runs necessary for a total ramp of $\Delta v_p= 300~\mathrm{m/s}$ took approximately 1400 processor-hours on a quad Opteron-64 processor machine using the Continuous Hugoniot Method.  A conservative estimate for the same work using conventional methods puts the processing time at approximately 20 processor-years on the same equipment.  Even on a massively parallel supercomputer, the computation time is extremely large.

\section{Conclusions}

This paper presents and validates the Continuous Hugoniot Method by direct comparison of its results with published data and with the results of conventional shock generation methods.  The method was applied to the very well studied Lennard-Jonesium test material and to the more realistic MEAM potential for tin.

We found very good agreement in each of our tests of the method.  We confirmed that the loading path of the Continuous Hugoniot Method in Lennard-Jones followed the published Hugoniot fit for both increasing and decreasing velocity ramps between $v_p = 1000\,\,\mathrm{m/s} = 0.75\,\,C_o$ and $v_p = 2000\,\,\mathrm{m/s} = 1.5\,\,C_o$.  In the range where the fit was not applicable, we were able to compare to conventional simulations methods with equally good agreement.  By comparing the results of the increasing and decreasing velocity ramps, we found that the system did exhibit some residual rate dependence, but was not hysteretic.  We found that the method also did relatively well in predicting the Hugoniot in the elastic-plastic regime.

Comparison of density profiles produced by our method showed very good agreement with conventional runs.  However, we found that the average density within our reduced system is low, on average, by approximately 2\% compared to the final density predicted in the Hugoniot relations.  Larger systems would give better agreement.  However, although not converged to the values predicted for far behind the shock front, our density profiles do make reliable predictions near the front.

Comparison of temperature profiles, particle snapshots and radial distribution functions at the final state of the Continuous Hugoniot Method ramp also demonstrated good agreement with conventional shock methods.

Application of the Continuous Hugoniot Method to the study of shock melting in tin clearly demonstrated its efficiency and effectiveness.  We found that tin both melts and comes to equilibrium within just a few nanometers of the shock front.  Mean square displacements versus time showed true liquid diffusion properties in the melt.

We presented a Hugoniot plot describing the shock response in tin and compared it with experimental data.  We found that the Hugoniot values for $U_S$ versus $v_p$ predicted by MEAM were approximately 6\% above those found in experiment.  However, differences between the experimental parameters and the simulation runs could account for such a discrepancy.  We plotted the temperature profiles of tin as a function of the shock strength parameter, $v_p$ and noted an anomalous transition at high driving velocity in which the temperature peak migrates abruptly forward in the shock profile.

The Continuous Hugoniot Method was applied to shocks in the two-shock elastic-plastic regime.  The transition from the strong shock regime was shown to be continuous and consistent with a transcritical bifurcation with order parameter equal to the relative velocity of the shock and precursor waves.  The fluctuations in the order parameter grew as the system approached onset, as expected. 

Finally, we were able to make these measurements with greatly reduced computational expenditure over conventional methods.  These savings proved critical when the method was applied to the more realistic and computationally costly MEAM potential.

\begin{acknowledgments}
The authors would like to thank Todd Ditmire, Stephan Bless and Will Grigsby for useful conversations. This work was supported by the National Science Foundation under DMR-0401766, and DMR-0101030 and by the U.S. Dept. of Energy, National Nuclear Security Administration under Contract DE-FC52-03NA00156.  Texas Applied Computing Center.
\end{acknowledgments}

\appendix

\section{MEAM Potential}
The total energy given by MEAM is

\begin{equation}
E = \sum_i \left(F(\bar \rho_i) + \frac{1}{2}\sum_{j \ne i} \phi(R_{ij})\right)
\label{e:meam_energy}
\end{equation}
where the sums are over particle indices; $F$ is the embedding energy as a function of $\bar\rho_i$, the background electron density at the site of the $i^{th}$ particle; and $\phi$ is the two-body interaction between particles $i$ and $j$ as a function of $R_{ij}$, their separation distance.

The embedding function, $F$, has the form
\begin{equation}
F(\bar \rho_i) = A\,E_c\,\bar\rho_i \ln \bar\rho_i
\end{equation}
where $A$ is a free parameter and $E_{c}$ is the cohesive energy.  The background electron density for tin is calculated at each occupied site by
\begin{equation}
\bar\rho = \rho^{(0)} \frac{2}{1+e^{-\Gamma}}
\end{equation}
where $\rho^{(0)}$ is the partial electron density associated with the spherically symmetric s orbital contributions from surrounding atoms.

$\Gamma$ is a weighted sum of the non-spherically symmetric partial electron densities associated with the p, f and g orbitals.  $\Gamma$ is positive definite for tin.

\begin{equation}
\Gamma = \sum_{h=1}^3t^{(h)}\left(\frac{\rho^{(h)}}{\rho^{(0)}}\right)^2
\end{equation}
$t^{(1)}$, $t^{(2)}$, and $t^{(3)}$ are parameters indicating the relative importance of each orbital, and the higher partial electron densities.

The second term from the MEAM total energy, Equation \ref{e:meam_energy}, is the two-body interaction term
\begin{equation}
\phi(R) = \frac{2}{Z}\left\{E^u(R) - F(\bar\rho^{\,0}(R))\right\}
\end{equation}
where $Z$ is the number of nearest neighbors in the reference structure and $F(\bar\rho^{\,0}(R))$ is the embedding energy of the reference structure background electron density $\bar\rho^{\,0}$.  $E^u(R)$ is the energy per atom in the reference structure as a function of the nearest-neighbor distance $R$.

In lieu of a potential cutoff distance, MEAM implements a many-body screening.  Thus the effective cutoff distance depends on the local environment.  Dense structures have short cutoffs; and open sparse structures, where there is little screening, can have long-range interactions.

The screening function $0 \le \zeta_{ik} \le 1$ multiplies the electron densities and the pair potential.  The total screening function is the product of screening terms for all particles $j$ which reside between the i and j particles.  The degree of screening is determined by
\begin{equation}
\zeta_{ik} = \prod_{j\ne i,k} S_{ijk}
\end{equation}
where $S_{ijk}$ is the screening effect of j between the $i^{th}$ and $j^{th}$ particles.  We use the simple algebraic form \cite{baskes.97} for the screening term.

We take our MEAM parameters from Ravelo and Baskes \cite{ravelo.97} without modification.

\bibliographystyle{aipproc}
\bibliography{../../bibtex/JMDL.bib}

\begin{thebibliography}{34}
\expandafter\ifx\csname natexlab\endcsname\relax\def\natexlab#1{#1}\fi
\providecommand{\enquote}[1]{``#1''}
\expandafter\ifx\csname url\endcsname\relax
  \def\url#1{\texttt{#1}}\fi
\expandafter\ifx\csname urlprefix\endcsname\relax\def\urlprefix{URL }\fi

\bibitem[Loveridge-Smith and et.al.(2001)]{loveridge-smith.01}
Loveridge-Smith, A., and et.al., \emph{Physical Review Letters}, \textbf{86},
  2349--2352 (2001).

\bibitem[Maillet et~al.(2000)]{maillet.00}
Maillet, J.~B., Mareschal, M., Soulard, L., Ravelo, R., Lomdahl, P.~S.,
  Germann, T.~C., and Holian, B.~L., \emph{Physical Review E}, \textbf{63},
  16121 (2000).

\bibitem[Reed et~al.(2002)]{reed.02}
Reed, E.~J., Joannopoulos, J.~D., and Fried, L.~E., \enquote{Hugoniot
  Constraint Molecular Dynamics Study of a Transformation to a Metastable phase
  in Shocked Silicon,} in \emph{Shock Compression of Comdensed Matter - 2001},
  edited by M.~D. Furnish, N.~N. Thadhani, and Y.~Horie, American Institute of
  Physics, AIP Proceedings, 2002, pp. 343--346.

\bibitem[Reed et~al.(2003)]{reed.03}
Reed, E.~J., Fried, L.~E., and Joannopoulos, J.~D., \emph{Physical Review
  Letters}, \textbf{90}, 235503 (2003).

\bibitem[Zhakhovskii et~al.(1999)]{zhakhovskii.99}
Zhakhovskii, V.~V., Zybin, S.~V., Nishihara, K., and Anisimov, S.~I.,
  \emph{Physical Review Letters}, \textbf{83}, 1175--1178 (1999).

\bibitem[Ahrens(1992)]{ahrens.92}
Ahrens, T.~J., \emph{High-Pressure Shock Compression in Solids},
  Springer-Verlag, 1992, chap. Equation of State, pp. 75--114.

\bibitem[Graham(1993)]{graham.92}
Graham, R.~A., \emph{High-Pressure Shock Compression of Solids},
  Springer-Verlag, 1993, chap. Introduction to High-Pressure Shock Compression
  of Solids, pp. 1--6.

\bibitem[Rosenberg et~al.(2001)]{rosenberg.01}
Rosenberg, Z., Bourne, N.~K., III, G. T.~G., and Millett, J. C.~F., \enquote{On
  the Measurement of Shear-strength in Quasi-isentropic Loading,} in
  \emph{Shock Compression of Condensed Matter}, 2001.

\bibitem[Holian(1988)]{holian.88}
Holian, B.~L., \enquote{Modeling Shockwave deformation via Molecular Dynamics,}
  in \emph{Shock Waves in Condensed Matter 1987}, edited by S.~Schmidt and
  N.~Holmes, Elsevier Science Publishers, 1988, pp. 185--190.

\bibitem[Paskin et~al.(1977)]{paskin.77}
Paskin, A., Gohar, A., and Dienes, G.~J., \emph{Journal of Physics C},
  \textbf{10}, L563--L566 (1977).

\bibitem[Holian(1995)]{holian.95}
Holian, B.~L., \emph{Shock Waves}, \textbf{5}, 149--157 (1995).

\bibitem[Belonoshko(1997)]{belonoshko.97}
Belonoshko, A.~B., \emph{Science}, \textbf{275}, 955--957 (1997).

\bibitem[Holian and Lomdahl(1998)]{holian.98}
Holian, B.~L., and Lomdahl, P.~S., \emph{Science}, \textbf{280}, 2085--2088
  (1998).

\bibitem[Holian and Straub(1979)]{holian.79}
Holian, B.~L., and Straub, G.~K., \emph{Physical Review Letters}, \textbf{43},
  1598--1600 (1979).

\bibitem[Lennard-Jones(1932)]{lennard-jones.32}
Lennard-Jones, J.~E., \emph{Transactions of the Faraday Society}, \textbf{28},
  333--359 (1932).

\bibitem[Holian et~al.(1991)]{holian.91}
Holian, B.~L., Voter, A.~F., Wagner, N.~J., Ravelo, R.~J., Chen, S.~P., Hoover,
  W.~G., Hoover, C.~G., Hammerberg, J.~E., and Dontje, T.~D., \emph{Physical
  Review A}, \textbf{43}, 2655--2661 (1991).

\bibitem[Germann et~al.(2000)]{germann.00}
Germann, T.~C., Holian, B.~L., and Lomdahl, P.~S., \emph{Physical Review
  Letters}, \textbf{84}, 5351--5354 (2000).

\bibitem[Rapaport(1995)]{rapaport.95}
Rapaport, D.~C., \emph{The Art of Molecular Dynamics Simulation}, Cambridge
  University Press, 1995.

\bibitem[Mabire and Hereil(2000)]{mabire.00}
Mabire, C., and Hereil, P.~L., \emph{Journal de Physique IV}, \textbf{10},
  749--754 (2000).

\bibitem[Cherne et~al.(2004)]{cherne.04}
Cherne, F.~J., Baskes, M.~I., Germann, T.~C., Ravelo, R.~J., and Kadau, K.,
  \enquote{Shock Hugoniot and Melt Curve for a Modified Embedded Atom Method
  Model of Gallium,} in \emph{Shock Compression of Condensed Matter - 2003},
  edited by M.~D. Furnish, Y.~M. Gupta, and J.~W. Forbes, American Physical
  Society, American Institute of Physics, 2004, vol. 706 of \emph{SCCM}.

\bibitem[Ravelo and Baskes(1997)]{ravelo.97}
Ravelo, R., and Baskes, M., \emph{Physical Review Letters}, \textbf{79},
  2482--2485 (1997).

\bibitem[Baskes(1992)]{baskes.92}
Baskes, M.~I., \emph{Physical Review B}, \textbf{46}, 2727--2742 (1992).

\bibitem[Daw and Baskes(1984)]{daw.84}
Daw, M.~S., and Baskes, M.~I., \emph{Physical Review B}, \textbf{29},
  6443--6453 (1984).

\bibitem[Baskes et~al.(1994)]{baskes.94}
Baskes, M.~I., Angelo, J.~E., and Bisson, C.~L., \emph{Modelling and Simulation
  in Material Science and Engineering}, \textbf{2}, 505--518 (1994).

\bibitem[Baskes(1997)]{baskes.97}
Baskes, M.~I., \emph{Materials Chemistry and Physics}, \textbf{50}, 152--158
  (1997).

\bibitem[Wyckoff(1924)]{wyckoff.24}
Wyckoff, R. W.~G., \emph{Structure of Crystals}, 1924.

\bibitem[Cahoon(2003)]{cahoon.03}
Cahoon, J.~R., \emph{Metallurgical and Materials Transactions A}, \textbf{34A},
  882--883 (2003).

\bibitem[Araki et~al.(1996)]{araki.96}
Araki, H., Minamino, Y., Yamane, T., Nakatsuka, T., and Miyamoto, Y.,
  \emph{Metallurgical and Materials Transactions A}, \textbf{27A}, 1807--1814
  (1996).

\bibitem[Reichl(1998)]{reichl.98}
Reichl, L.~E., \emph{A Modern Course in Statistical Physics}, John Wiley \&
  Sons, 1998.

\bibitem[Marsh(1980)]{marsh.80}
Marsh, S.~P., Lasl shock hugoniot data, Univ. of California Press, Berkeley
  (1980).

\bibitem[http://teos.ficp.ac.ru/rusbank(????)]{russianweb}
http://teos.ficp.ac.ru/rusbank, Russian shock database (????).

\bibitem[Hardy(1982)]{hardy.82}
Hardy, R.~J., \emph{Journal of Chemical Physics}, \textbf{76}, 622--628 (1982).

\bibitem[Gilvarry(1956)]{gilvarry.56}
Gilvarry, J.~J., \emph{Physical Review}, \textbf{102}, 308--316 (1956).

\bibitem[Rice et~al.(1958)]{rice.58}
Rice, M.~H., McQueen, R.~G., and Walsh, J.~M., \enquote{Solid State Physics,}
  in \emph{Compression of Solids by Strong Shock Waves}, 1958, vol.~6 of
  \emph{Solid State Physics}.

\end{thebibliography}

\end{document}